\documentclass[12pt]{article}
\usepackage{amscd,amssymb,amsmath,latexsym,enumerate}
\usepackage[mathscr]{euscript}
\usepackage{epsfig}
\usepackage{textcomp}
\usepackage{amsmath}
\usepackage{amssymb}
\usepackage{verbatim}

\title{Persistence of spin edge currents \\ in disordered quantum spin Hall systems}
\vspace{.5cm}

\author{Hermann Schulz-Baldes
\\
{\small Department Mathematik, Universit\"at Erlangen-N\"urnberg, D-91058
Erlangen, Germany} 
\vspace{.2cm}
}

\date{ }

\newcommand{\CM}{{\mathbb C}}

\newcommand{\NM}{{\mathbb N}}

\newcommand{\RM}{{\mathbb R}}

\newcommand{\ZM}{{\mathbb Z}}

\newcommand{\Oo}{{\cal O}}

\newcommand{\Tt}{{\cal T}}
\newcommand{\Hh}{{\cal H}}

\newcommand{\EE}{{\bf E}}                      
\newcommand{\PP}{{\bf P}}                      

\newcommand{\dis}{{\mbox{\rm\tiny dis}}}
\newcommand{\hc}{{\mbox{\rm\tiny hc}}}
\newcommand{\Ra}{{\mbox{\rm\tiny Ra}}}
\newcommand{\Ze}{{\mbox{\rm\tiny Ze}}}
\newcommand{\SO}{{\mbox{\rm\tiny SO}}}
\newcommand{\Ch}{\mbox{\rm Ch}}
\newcommand{\SCh}{\mbox{\rm SCh}}
\newcommand{\Tr}{\mbox{\rm Tr}}
\newcommand{\one}{{\bf 1}}

\def\XXint#1#2#3{{\setbox0=\hbox{$#1{#2#3}{\int}$}
     \vcenter{\hbox{$#2#3$}}\kern-.5\wd0}}

\begin{document}

\maketitle

\begin{abstract} 
For a disordered two-dimensional model of a topological insulator (such as a Kane-Mele model with disordered potential) with small coupling of spin invariance and time-reversal symmetry breaking terms (such as a Rashba spin-orbit coupling and a Zeeman term), it is proved that the spin edge currents persist provided there is a spectral gap and the spin Chern numbers are well-defined and non-trivial. This are sufficient conditions for being in the quantum spin Hall phase. The result materializes the general philosophy that topological insulators are topologically non-trivial bulk systems with persistent edge or surface currents. 
\end{abstract}

\section{Resum\'e and discussion}

A quantum spin Hall system is a two-dimensional solid state phase resulting from a strong spin-orbit interaction.  Kane and Mele proposed a tight-binding approximation for graphene as the first toy model in this phase \cite{KM1,KM2}. It has non-trivial bulk topology going along with helical edge states which carry spin currents that are not exposed to Anderson localization by a disordered potential.  Furthermore, these systems exhibit the intrinsic spin Hall effect (a dissipationless bulk spin current as a reaction to an applied electric field even in absence of magnetic impurities) as well as the inverse spin Hall effect (an electric current resulting from a gradient of the Zeeman field) \cite{ERH,SWSH,Mur}. The quantum spin Hall system is the prime example of so-called topological insulators \cite{SRFL,HK}. Mainly based on numerics and analogies with quantum Hall systems, it is usually argued that non-trivial invariants of the bulk lead to non-trivial spin edge currents \cite{SRFL,KM2,HK}. A first rigorous proof of such a connection for periodic systems was given in our prior work \cite{ASV}. Independently, Graf and Porta provide an alternative proof applying to more general situations \cite{GP}. While interesting, this is definitely not yet satisfactory because disorder is an essential ingredient to quantum spin Hall systems and topological insulators in general. The main result of this work examines the stability of the spin edge currents under disordered and time-reversal symmetry breaking perturbations:

\vspace{.2cm}

\noindent {\bf Theorem} {\it For a disordered two-dimensional one-particle tight-binding model with spectral gap, approximate spin conservation and small Zeeman term, the spin edge currents do not vanish provided the spin Chern numbers are non-trivial.}

\vspace{.2cm}

The concrete model, explicit hypothesis and detailed result are described below. The argument is of perturbative nature around $s^z$-invariant Hamiltonians (here $s^z$ is the $z$-component of the spin operator). Such Hamiltonians can be decomposed into a direct sum of Hamiltonians acting on the eigenspaces of $s^z$. In presence of a gap, the Fermi projection of each such operator has a well-defined Chern number which is then called the spin Chern number of the total Hamiltonian \cite{SWSH,Pro}. Non-trivial spin Chern numbers lead to quantized edge currents for each subsystem by the argument for disordered quantum Hall systems \cite{KRS,EG}. If the spin-weighted sum of these currents does not vanish, it only remains to control by perturbation theory the effect of $s^z$-invariance breaking terms. This is done in a quantitative manner in this paper. This may seem as a small addendum to prior works which, moreover, only faces minor technical difficulties, but it is nevertheless important because it is the first rigorous result supporting the conception of topological insulators widely accepted in the physics community, notably as disordered quantum systems with non-trivial bulk topology which implies edge or surface currents. 

\vspace{.2cm}

A further somewhat surprising implication of the theorem is that spin edge currents also do not vanish when a Zeeman term is added which opens a gap in the edge state spectrum (and, moreover, breaks time-reversal invariance). The claim is that nevertheless the remaining edge states away from the gap in the edge spectrum carry spin current.  Until recently it was firmly believed (see all the references cited above) that the quantum spin Hall phase is tied to a non-trivial $\ZM_2$-invariant defined only for time-reversal invariant systems with odd spin. However, several newer works \cite{SP,YXS,XSX} indicate that rather non-trivial spin Chern numbers summing up to zero are the main and defining characteristic of the novel phase. In particular, time-reversal invariance is not a crucial requisite for the quantum spin Hall phase which persists, {\it e.g.}, even in presence of a Zeeman term \cite{YXS}. If the spin Chern numbers are non-trivial, but their sum does not vanish, one is rather in a phase of quantum anomalous Hall effect. The definition of spin Chern numbers (and thus the quantum spin Hall phase) is not restricted to time-reversal invariant systems and readily extends to disordered systems as pointed out by Prodan \cite{Pro} because all that is needed is a gap in the spectrum of $P s^z P$ (here $P$ is the Fermi projection). Actually, this gap may remain open even though the gap of the Hamiltonian closes and becomes a mobility gap. In such a mobility gap, it is nevertheless possible to have well-defined edge currents if an adequate procedure for their definition is used (see \cite{EGS} and the comment at the end of the paper). It is thus possible that the statement of the above theorem also holds if the spectral gap condition is replaced by a mobility gap condition. In laboratory systems \cite{KWB,BRB} there seems to be a true gap though and the role of the mobility gap is not as central as in the quantum Hall effect where no quantization would be possible without it. In view of all this, the spectral gap of $P s^z P$ combined with non-trivial spin Chern numbers adding up to zero are the most adequate indicator of the quantum spin Hall phase. 

\vspace{.2cm}

There is an important subtlety concerning the physical interpretation of the main theorem which is linked to the choice of the spin current operator. In this paper, the operator $J=\frac{1}{2}(s^z\,\imath[H,X]+\imath[H,X]\,s^z)$ is used where $X$ is the position operator and $H$ is the Hamiltonian. This is the symmetrized form of "spin times velocity", a definition that was and still is widely used (see the review \cite{ERH}). It could also be termed a "spin-polarized current operator". However, it is also possible to use the "time-derivative of the spin position" $G=\imath[H,s^zX]$ as the spin current operator. This choice was advocated in \cite{SZXN} because it leads to a conservation equation for spin as well as conjugate operators in the thermodynamical sense so that Onsager relations immediately hold. Furthermore, if $G$ is used for the calculation of the spin Hall conductivity, a Streda-like formula for the spin Hall conductivity can be derived \cite{Mur}. The difference $G-J=\frac{1}{2}(X\,\imath[H,s^z]+\imath[H,s^z]\,X)$ does not vanish if the spin is not globally conserved (namely $[H,s^z]\not=0$), a situation that is relevant for quantum spin Hall models including a Rashba term.  From a mathematical point of view, $J$ has the pleasant feature of being bounded and covariant. Both these properties fail to hold for $G$ which leads to technical difficulties, but also questions the physical relevance of $G$. On the other hand, it has been pointed out \cite{Ras,SZXN} that  $J$ not only describes transport spin currents, but also stationary circulating spin currents. Such currents can pictorially be represented by a one-dimensional oscillatory motion for which velocity and spin are both reverted at the turning points. In particular, such spin currents can also be carried by localized states and therefore the theorem does not exclude Anderson localization of the edge states. Such Anderson localization is, in particular, expected for the quasi-one-dimensional system of edge states if the perturbation breaks time-reversal symmetry. If one uses an effective description of the edge states by quasi-one-dimensional random Dirac operators, localization can indeed be proved for time-reversal symmetry breaking random perturbations under adequate coupling hypothesis (leading to positive Lyapunov exponents), while for time-reversal symmetric random perturbations there is always a conducting channel (pair of vanishing Lyapunov exponents) for symmetry reasons if the number of edge channels is odd \cite{SS}.  Hence several scenarios for time-reversal-breaking perturbations are conceivable and compatible with the results of this paper: (i) all edge states localize (and the spin currents are only circulating), or (ii) all edge states remain delocalized, or (iii) there is is an Anderson transition in the edge spectrum. Because there are several indications that the bulk states are delocalized \cite{XSX} the author does expect delocalized edge states in this situation (even for weak time-reversal symmetry breaking terms). This agrees also with the recent experimental observation of spin-polarized edge currents \cite{BRB}, which complement earlier measurements of spin accumulation at the edge of the sample \cite{KWB}. What can be said for sure is that the spin edge currents exhibited in the theorem are signatures of non-trivial bulk topology similar as edge charge currents are an indication for non-trivial orbital magnetization of the bulk \cite{ST}. 

\vspace{.2cm}

The remainder of the paper mainly contains the statement and proof of the result described above, except for one addendum. Because of their central role for the quantum spin Hall effect, it is of crucial importance to have efficient means to calculate the spin Chern numbers of a given model. Many techniques for this purpose are spread in the literature. Here an efficient algorithm from \cite{ASV} is sketched  at the end of Section~\ref{sec-spinChern} which allows to calculate the spin Chern numbers for periodic systems using only the spectral analysis of finite size transfer matrices. 

\vspace{.2cm}

\noindent {\bf Acknowledgements:} The author thanks Julio Cesar Avila and Carlos Villegas-Blas for the collaboration \cite{ASV} which led to the natural question examined in this paper, Emil Prodan and the referees for correspondence about the first versions of this work, and Gian-Michele Graf for several interesting discussions on quantum Hall systems over the years, and topological insulators more recently. This work was supported by the DFG.

\vspace{.2cm}

\section{Models and hypothesis}

Let us consider a two-dimensional system of independent tight-binding Fermions with half-integer spin $s\in\frac{1}{2}\,\NM\setminus\NM$ on the square lattice $\ZM^2$ where over each point of $\ZM^2$ there are $R\in\NM$ internal degrees of freedom. Hence the Hilbert space is $\ell^2(\ZM^2)\otimes\CM^R\otimes \CM^r$ where $r=2s+1$, but we also set $L=Rr$ and write $\CM^L=\CM^R\otimes \CM^r$. This setting allows to study particles on square, triangular and honeycomb lattice (see \cite{ASV}, although this can probably also be found elsewhere). 
%
%
The Hamiltonian is a strongly continuous family $(H_\omega)_{\omega\in\Omega}$ of self-adjoint finite difference operators on $\ell^2(\ZM^2)\otimes \CM^L$ indexed by a variable $\omega$, called the disorder configuration, taken from a compact dynamical system $(\Omega,T,\ZM^2)$ and satisfying the covariance relation
$$
H_{T_a\omega}\;=\;U_a \,H_\omega\,U_{a}^*\;,
\qquad
a\in\ZM^2
\;,
$$
where $U_a$ are the magnetic translations defined 
$$
(U_a\phi)_n\;=\;
e^{\frac{\imath}{2} B \,n\wedge a}\;\phi_{n-a}\;,
\qquad
n\wedge a=n_1a_2-n_2a_1
\;.
$$  
We also simply write $H=(H_\omega)_{\omega\in\Omega}$ (and this is then identified with an element of the C$^*$-algebra of homogeneous observables used in \cite{BES,KRS}, but this will not be used here). The Hamiltonians $H_\omega$ are supposed to have matrix (hopping) elements of at most uniformly bounded finite range. The model is completed by fixing an invariant and ergodic probability $\PP$ on $\Omega$. The main assumptions of the paper are the following:

\vspace{.2cm}

\noindent {\bf Gap hypothesis:} {\it There is an energy $E_g$ lying in a gap of the almost-sure spectrum of $H=(H_\omega)_{\omega\in\Omega}$. }

\vspace{.2cm}

\noindent {\bf Approximate spin conservation:} {\it The norm of the commutator $[H_\omega,s^z]$ is uniformly bounded in norm by some sufficiently small constant $C_{\mbox{\tiny\rm s}}<1$ smaller than the size of the gap.}

\vspace{.2cm}

One consequence of these hypothesis is that there exists a homotopy $H(\lambda)=(H_\omega(\lambda))_{\omega\in\Omega}$ of covariant Hamiltonians between $H(1)=H$ and $H(0)$ commuting with $s^z$ such that the gap $E_g(\lambda)$ associated to $E_g$ remains open along the homotopy (as $C_{\mbox{\tiny\rm s}}$ is sufficiently small). It can be chosen as
\begin{equation}
\label{eq-homotopy}
H(\lambda)\;=\;H\;+\;\frac{1-\lambda}{2}\,(s^z\,H\,s^z-H)
\;=\;H\;+\;\frac{1-\lambda}{2}\,[s^z,H]\,s^z
\;,
\qquad
\lambda\in[0,1]
\;.
\end{equation}
By construction one has $[H(0),s^z]=0$ so that these two self-adjoint operators can be simultaneously diagonalized so that
\begin{equation}
\label{eq-Hdecomp}
H(0)
\;=\;
H_{-s}(0)\oplus\ldots\oplus  H_{s}(0)
\;,
\end{equation}
where the direct sum decomposition is w.r.t. the eigenspaces of $s^z$, that is $\ell^2(\ZM^2)\otimes\CM^L=\oplus_{l=-s}^s\Hh_l$ where $\Hh_l=\ell^2(\ZM^2)\otimes\CM^R$. This means that the Hamiltonians decomposes into $r=2s+1$ separate Hamiltonians, each of which can be studied as a quantum Hall system along the lines of \cite{BES,KRS}.  This will be heavily used further below.

\vspace{.2cm}

Let us provide a concrete situation of physical interest where both the above hypothesis are satisfied, namely the Kane-Mele model on a honeycomb lattice with spin-orbit interactions. In this situation, there are two internal degrees of freedom ($R=2$) resulting from the bipartite structure of the lattice. The Hamiltonian on $\ell^2(\ZM^2)\otimes \CM^{2r}$ is now of the form
$$
H_{\omega}
\;=\;
H_\hc\,+\,\lambda_\SO\,H_\SO\,+\,\lambda_\Ra\,H_\Ra\,+\,\lambda_\Ze\,H_\Ze\,+\,
\lambda_\dis \sum_{n\in\ZM^2}V_n\,|n\rangle\langle n|
\;.
$$
Here $H_\hc$ is the discrete Laplacian on the honeycomb lattice, $H_\SO$ is the second nearest neighbor spin orbit coupled hopping, $H_\Ra$ is the next nearest neighbor Rashba spin orbit hopping (all possibly with magnetic field), $H_\Ze$ is a Zeeman term, and finally the $V_n$ are independent random $2r\times 2r$ matrices which are supposed to be real and commute with $s^z$ (typically they are diagonal with two random entries describing random potential values on two neighboring sites of the honeycomb lattice). A detailed description of the Hamiltonian (except $H_\Ze$) is given in \cite{KM2,ASV}. The main features of importance here are that $[H_\SO,s^z]=0$, $[H_\Ra,s^z]\not =0$ and $[H_\Ze,s^z]\not =0$ so that the Rashba and Zeeman terms are the only one breaking conservation of the $s^z$ component. It is known \cite{KM2} that for $s=\frac{1}{2}$, $\lambda_\Ra=\lambda_\Ze=\lambda_\dis=0$ and $\lambda_\SO>0$ the energy $E_g=0$ lies in a gap of the spectrum and there is one band above and one below $E_g=0$. For sufficiently small $\lambda_\Ra$, $\lambda_\Ze$ and $\lambda_\dis$ (smaller, but possibly of the order of $\lambda_\SO$) the above Gap hypothesis as well as the Approximate spin conservation both remain valid. Moreover, as discussed once again below, the model is a non-trivial topological insulator in this regime because the spin Chern numbers defined below do not vanish. Let us also point out that by multiplying $\lambda_\Ra$ by $\lambda\in [0,1]$, one obtains another homotopy $H_{\omega}(\lambda)$ from $H_\omega(1)=H_{\omega}$ to $H_{\omega}(0)$ which then commutes with $s^z$. This homotopy does not coincide with \eqref{eq-homotopy}, but may just as well be used below.

\section{Spin Chern numbers}
\label{sec-spinChern}

The spectrum of the spin operator $s^z$ is $\{-s,-s+1,\ldots,s-1,s\}$, and each of these eigenvalues is infinitely degenerate. Following Prodan \cite{Pro}, let us now consider the self-adjoint operator $P(\lambda) s^z P(\lambda)$ where $P(\lambda)=(P_\omega(\lambda))_{\omega\in\Omega}$ denotes the spectral projection of $H(\lambda)=(H_\omega(\lambda))_{\omega\in\Omega}$ given in \eqref{eq-homotopy} below the gap $E_g(\lambda)$.
For $\lambda=0$, one has $[P(0), s^z]=0$ so that 
\begin{equation}
\label{eq-Spindiag}
s^z\,P_l(0)\;=\;l\,P_l(0)
\;.
\end{equation}
In particular, the spectrum of $P(0) s^z P(0)$ is still $\{-s,\ldots,s\}\cap\{0\}$ with each level infinitely degenerate. As $\lambda$ increases, this degeneracy is lifted leading to $r+1=2s+2$ spectral islands, but at least if $C_{\mbox{\rm\tiny s}}$ is sufficiently small, there remain spectral gaps between these spectral islands. Hence it is possible to choose positively oriented contours $\gamma_l$ around each one of them and define a covariant family of orthogonal projections $P_{l}(\lambda)=(P_{\omega,l}(\lambda))_{\omega\in\Omega}$ by
$$
P_{\omega,l}(\lambda)
\;=\;
\oint_{\gamma_l}\frac{d\xi}{2\pi\imath}\;
\bigl(\xi-P_\omega(\lambda) s^z P_\omega(\lambda)\bigr)^{-1}
\;,
\qquad
l=-s,\ldots,s
\;.
$$
Clearly, one has
$$
P(\lambda)\;=\;\sum_{l=-s}^s\,P_{l}(\lambda)\;.
$$
As $P_\omega(\lambda)$ has exponentially decreasing matrix elements due to the presence of the spectral gap, a standard argument (of Combes-Thomas type) shows that also $P_{\omega,l}(\lambda)$ has exponentially decreasing matrix elements. Therefore the Chern number of $P_{l}(\lambda)$ in the sense of \cite{BES,ASS} is well-defined and an integer:
$$
\Ch(P_{l}(\lambda))
\;=\;
\frac{1}{2\pi\imath}\;
\EE_\PP\;\Tr_L\,\langle 0|
P_{\omega,l}(\lambda)
[[X_1,P_{\omega,l}(\lambda)],[X_2,P_{\omega,l}(\lambda)]]|0\rangle
\;.
$$
Here $\Tr_L$ denotes the trace over $L\times L$ matrices. As these integers are intrinsic to $P(\lambda)$ and its interplay with the spin operator $s^z$, they are called the spin Chern numbers of $P(\lambda)$ and denoted by
$$
\SCh_l(P(\lambda))
\;=\;
\Ch(P_{l}(\lambda))
\;,
\qquad
l=-s,\ldots,s
\;.
$$
By the additivity of Chern numbers, one has
\begin{equation}
\label{eq-Chernsum}
\Ch(P(\lambda))
\;=\;
\sum_{l=-s}^s\;\SCh_l(P(\lambda))
\;.
\end{equation}
Because $\lambda\in[0,1]\mapsto P(\lambda)$ is a norm continuous path of projections, the spin Chern numbers are independent of $\lambda$ due to the homotopy invariance of the Chern numbers \cite{BES}. For time-reversal invariant systems, the Chern number $\Ch(P(\lambda))$ always vanishes because $\Ch(P(\lambda))=\Ch(\Theta P(\lambda)\Theta^{-1})=-\Ch(P(\lambda))$ if $\Theta$ is the time-reversal operator. This does not mean that all of the spin Chern numbers vanish though.  It is worth mentioning that the spin Chern numbers are also well-defined if the gap assumption is replaced by a dynamical (Anderson) localization condition, but this is not used below where only edge states in the gap are studied.

\vspace{.2cm}

Of course, it is of crucial importance to be able to calculate the spin Chern numbers in a given model in order to determine whether or not one is in a topologically non-trivial quantum spin Hall phase. Due to the homotopy invariance properties of the spin Chern numbers discussed above, it is often sufficient to do this for a $s^z$-invariant model without disorder.  For simplicity let us also assume that  the magnetic field vanishes (rational magnetic fields can be dealt with similarly, and small magnetic fields may again be turned off by a homotopy). Then the Hamiltonian decomposes into a direct sum \eqref{eq-Hdecomp} of periodic operators. If the hopping in the original model is nearest neighbor, then one such operator $H_l$ acting on $\ell^2(\ZM^2)\otimes \CM^R$ is of the following form:
\begin{equation}
\label{eq-Jacobiform}
H_l
\; = \;
T_1S_1^*+ T_1^*S_1+
T_2S_2^*+T_2^*S_2
+T_3S_3^*+T_3^*S_3
+T_4S_4^*+T_4^*S_4
+W
\;.
\end{equation}
Here $S_1$ and $S_2$ are the shift operators on $\ell^2(\ZM^2)$ and $S_3=S_1^*S_2$ and $S_4=S_1S_2$, and furthermore the $T_j$, $j=1,2,3$, and $W=W^*$ are self-adjoint $R\times R$ matrices. If the hopping is not nearest neighbor, but finite range, this form can again be achieved by enlarging the number $R$ of internal degrees of freedom (or in other words, enlarging the unit cell). As shown explicitly in \cite{ASV}, operators on quadratic, triangular and hexagonal lattice can be brought into the form \eqref{eq-Jacobiform} even if $T_4$. Here the term $T_4$ is added because it does not lead to further complications in the arguments and allows to treat further models. One now has to calculate the Chern number of the Fermi projection $P_l$ associated to $E_g$ of one such operator $H_l$. The following procedure to achieve this is explained and justified in detail in \cite{ASV}. First build the family in $k\in[-\pi,\pi]$ of $2R\times 2R$ transfer matrices
$$
{\Tt}(k)\;=\;
\begin{pmatrix}
(E_g\,{\bf 1}-(e^{\imath k} T_1^*+e^{-\imath k}T_1+W))(T_2+e^{\imath k}T_3+e^{-\imath k}T_4)^{-1}
& -\,(T_2+e^{\imath k}T_3+e^{-\imath k}T_4)^*
\\
(T_2+e^{\imath k}T_3+e^{-\imath k}T_4)^{-1} & 0
\end{pmatrix}
\;,
$$
under condition that the inverse of $T_2+e^{\imath k}T_3+e^{-\imath k}T_4$ exists (this is required at least for almost all $k$). Because $E_g$ is in a gap of the spectrum of $H_l$, this transfer matrix is hyperbolic (no eigenvalues of modulus $1$) and due to its symplectic symmetry it has exactly $R$ eigenvalues inside of the unit disc and $R$ reflected ones outside. Determine (numerically if necessary) $R$ linearly independent generalized eigenvectors $\phi_m(k)\in\CM^{2R}$ associated to the eigenvalues inside the unit disc and build an $2R\times R$ matrix $\Phi(k)=(\phi_1(k),\ldots,\phi_R(k))$. Then set
$$
U(k)\;=\;\binom{1}{\imath}^*\Phi(k)\left( \binom{1}{-\imath}^*\Phi(k)\right)^{-1}
\;,
$$
where $1$ and $\pm\imath$ are multiplied by the $R\times R$ identity matrix and $\binom{1}{\pm\imath}^*=(1\;\mp \,\imath)$. The $R\times R$ matrix $U(k)$ (and, in particular, the appearing operator inverse) is well-defined and turns out to be unitary. The associated winding number is equal to the desired Chern number:
$$
\Ch(P_l)
\;=\;
\int^\pi_{-\pi}\frac{dk}{2\pi\imath}\;\ln\;\det(U(k))
\;.
$$
This procedure can easily be implemented numerically \cite{ASV}. For example, for the spin $s=\frac{1}{2}$ Kane-Mele model with $\lambda_\Ra=\lambda_\dis=0$ one finds that the two spin Chern numbers of the lower band are $1$ and $-1$. By homotopy, this remains true in the regime $\lambda_\SO>\max\{\lambda_\Ra,\lambda_\dis\}$.

\section{Spin edge currents}

In order to study edge currents, we now simply restrict all operators to a half-space. Let $\Pi:\ell^2(\ZM^2)\to\ell^2(\ZM\times\NM)$ denote the partial isometry from $\ell^2(\ZM^2)$ onto $\ell^2(\ZM\times\NM)$, namely $\Pi\Pi^*$ is the identity on $\ell^2(\ZM\times\NM)$ and $\Pi^*\Pi$ is the projection on $\ell^2(\ZM\times\NM)$ seen as subspace of $\ell^2(\ZM^2)$. Then $\widehat{H}(\lambda)=(\widehat{H}_\omega(\lambda))_{\omega\in\Omega}$ is defined by $\widehat{H}_\omega(\lambda)=\Pi H_\omega(\lambda)\Pi^*$. It is possible to include other covariant boundary conditions, but this leads to some further technicalities that are inessential. The family $\widehat{H}(\lambda)=(\widehat{H}_\omega)_{\omega\in\Omega}$ is still covariant in the $1$-direction, but not the $2$-direction. The observable for edge currents along the boundary is now the $1$-component $\widehat{J}_1(\lambda)$ of the spin current operator. It is given by the $1$-component of the velocity  $\dot{X}_1=\imath [\widehat{H}(\lambda),X_1]$ multiplied with the $z$-component of the spin operator, albeit in a symmetrized form:
$$
\widehat{J}_1(\lambda)
\;=\;
\frac{1}{2}
\left(s^z\;\imath [\widehat{H}(\lambda),X_1]
\;+\;
\imath [\widehat{H}(\lambda),X_1]\;s^z
\right)
\;.
$$
This operator is bounded and covariant, but as discussed in the introduction other definitions are possible. Now let $g(\widehat{H})$ be a non-negative operator defined by spectral calculus from a non-negative function $g$ on $\RM$ with $\int dE\,g(E)=1$ and compact support lying in a gap of $H$. The operator $g(\widehat{H})$ can be interpreted as a density matrix of edge states.  The edge current $j^{\mbox{\rm\tiny e}}(g,\lambda)$ is then defined by
\begin{equation}
\label{eq-edgecurrentdef}
j^{\mbox{\rm\tiny e}}(g,\lambda)
\;=\;
\widehat{\Tt}\bigl(\widehat{J}_1(\lambda)\,g(\widehat{H}(\lambda))\bigr)
\;,
\end{equation}
where $\widehat{\Tt}=\Tt_1\,\Tr_2$ is the trace per unit volume in the $1$-direction along the boundary and $\Tr_2$ the usual trace in the $2$-direction perpendicular to the boundary. For any operator family $\widehat{A}=(\widehat{A}_\omega)_{\omega\in\Omega}$ on $\ell^2(\ZM\times\NM)$ which is homogeneous in the $1$-direction, a formal definition of $\widehat{\Tt}$ is
\begin{equation}
\label{eq-Thatdef}
\widehat{\Tt}(\widehat{A}\,)
\;=\;
\EE_\PP\;\Tr_L\; \sum_{n_2\geq 0}\;\langle 0,n_2|\widehat{A}_\omega|0,n_2\rangle
\;.
\end{equation}
Further properties of $\widehat{\Tt}$ are collected in \cite{KRS} (where $\widehat{\Tt}$ does not contain the trace $\Tr_L$, but this leads to no changes).  In particular, $\widehat{\Tt}$ is cyclic and the $\widehat{\Tt}$-traceclass operators form a two-sided ideal in the algebra of covariant operators. As is obvious from this definition, not all covariant operators $\widehat{A}$ are traceclass w.r.t. $\widehat{\Tt}$ due to the sum over $n_2$ in \eqref{eq-Thatdef}. Thus one has to prove that \eqref{eq-edgecurrentdef} actually makes sense for sufficiently smooth functions $g$.  This is part of the following main result of the paper. 

\vspace{.2cm}

\noindent{\bf Theorem} {\it
Let $H=(H_\omega)_{\omega\in\Omega}$ satisfy the {\rm Gap hypothesis} and the {\rm Approximate spin conservation}. Then, if $g\in C^6(\RM)$ is any positive function of unit integral supported in the spectral gap associated to $E_g$ and $P$ is the Fermi projection of $H$ beneath this gap,
\begin{equation}
\label{eq-spinedgecur}
j^{\mbox{\rm\tiny e}}(g,\lambda)
\;=\;
\sum_{l=-s}^s\;l\;\SCh_l(P)
\;+\;
\lambda\,R(g,\lambda)
\;,
\end{equation}
where $R(g,\lambda)$ is a function satisfying for some constant $C$ 
$$
|R(g,\lambda)|
\;\leq\;C\,\|g\|_6\,\|[s^z,H]\|
\;.
$$
}

\vspace{.1cm}

As already discussed above, the result shows that the spin edge currents are quantized for $s^z$-invariant Hamiltonians and that they are non-vanishing if $\sum_{l=-s}^s\;l\;\SCh_l(P)\not = 0$ and 
both $\|[s^z,H]\|$ and $\|g\|_6$ are sufficiently small. In particular, the support of $g$ cannot be chosen arbitrarily small for a given value of $\|[s^z,H]\|$.

\section{Proof}

First of all, at $\lambda=0$ the Hamiltonian decomposes into a direct sum \eqref{eq-Hdecomp}. For each summand $H_l(0)$ acting on $\ell^2(\ZM^2)\otimes\CM^R$, the Chern number of the Fermi projection $P_l(0)$ below $E_g$ is $\Ch(P_l(0))=\SCh_l(P(0))$. Therefore the results of 
\cite{KRS,EG} (which also contain the $\widehat{\Tt}$-traceclass property of $g(\widehat{H}(0))$ that is shown once again below) imply
$$
\widehat{\Tt}\bigl(\imath[\widehat{H}_l(0),X_1]\,g(\widehat{H}_l(0))\bigr)
\;=\;
\SCh_l(P(0))
\;,
$$
where here $\widehat{\Tt}$ only contains a trace of $\CM^R$ and not $\CM^L$. Due to  \eqref{eq-Spindiag}, summing over $l\in\{-s,\ldots,s\}$ shows that \eqref{eq-spinedgecur} holds for $\lambda=0$. Therefore, only remains to develop a controlled perturbation theory in $\lambda$.

\vspace{.2cm}

Let us set $H_0=H(0)$ and $H_1=\frac{1}{2}[H,s^z]s^z$ so that according to \eqref{eq-homotopy} one has $H=H_0+\lambda H_1$ with $[H_0,s^z]=0$ and $\|H_1\|\leq \frac{s}{2}\|[H,s^z]\|$. The index $\omega$ will be suppressed in the following. The first step will then be to decompose
\begin{equation}
\label{eq-gdecomp}
g(\widehat{H}(\lambda))
\;=\;
g(\widehat{H}(0))
\;+\;\lambda\,G(\lambda)
\;,
\end{equation}
where $G(\lambda)$ is some remainder. As $\widehat{J}_1(\lambda)=\widehat{J}_1(0)+\frac{\lambda}{2}( s^z\,\imath [\widehat{H}_1,X_1]+\imath [\widehat{H}_1,X_1]\,s^z)$, reassembling  the terms shows
$$
j^{\mbox{\rm\tiny e}}(g,\lambda)
\;=\;
j^{\mbox{\rm\tiny e}}(g,0)\;+\;
\frac{\imath\,\lambda}{2}\;\widehat{\Tt}\bigl((s^z[\widehat{H}_1,X_1]+[\widehat{H}_1,X_1] \,s^z)\,g(\widehat{H}_0)\bigr)
\;+\;
\lambda\;\widehat{\Tt}\bigl(\widehat{J}_1(\lambda)G(\lambda)\bigr)
\;.
$$
The aim is thus to prove bounds on the last two summands. As in \cite{EG}, let us start from the Helffer-Sj\"ostrand formula for an arbitrary self-adjoint and bounded operator $H$:
\begin{equation}
\label{eq-HS}
g(H)
\;=\;
\frac{-1}{2\pi}\;
\int_{\RM^2}
dx\,dy\;
\partial_{\overline{z}}\widetilde{g}(x,y)\;(z-H)^{-1}\;,
\qquad
z=x+\imath y
\;,
\end{equation}
where, for some $N\geq 1$,
$$
\widetilde{g}(x,y)
\;=\;
\sum_{n=0}^N \;g^{(n)}(x)\;\frac{(\imath y)^n}{n!}\;\chi(y)\;,
$$
with some smooth, even, compactly supported function $\chi:(-1,1)\to [0,1]$ which is equal to $1$ on $[-\delta,\delta]$. The integral in \eqref{eq-HS} is norm convergent and the function $\widetilde{g}$, called a quasi-analytic extension of $g$, satisfies
$$
\partial_{\overline{z}}\widetilde{g}(x,y)
\; = \;
g^{(N+1)}(x)\;\frac{(\imath y)^N}{N!}\;\chi(y)
\,+\,\imath\,
\sum_{n=0}^N g^{(n)}(x)\;\frac{(\imath y)^n}{n!}\;\chi'(y)
\;,
$$
so that, in particular, uniformly in $x,y$, 
\begin{equation}
\label{eq-gtildebound}
|\partial_{\overline{z}}\widetilde{g}(x,y)|\;\leq \;C_0\,\|g\|_{N+1}\,|y|^{N}
\;.
\end{equation}
We use \eqref{eq-HS} to calculate $g(\widehat{H}(\lambda))$. Next let us replace the geometric resolvent identity
\begin{equation}
\label{eq-geomresol}
\frac{1}{z-\widehat{H}(\lambda)}
\;=\;
\Pi\,\frac{1}{z-H(\lambda)}\,\Pi^*
+\frac{1}{z-\widehat{H}(\lambda)}
\,(\widehat{H}(\lambda)\,\Pi^*-\Pi\, H(\lambda))\,\frac{1}{z-H(\lambda)}\,\Pi^*
\mbox{  .  }
\end{equation}
Replaced in \eqref{eq-HS}, the first summand gives $\Pi\, g(H(\lambda))\,\Pi^*$ which vanishes because the support of $g$ is contained in a gap of the spectrum of $H(\lambda)$. Thus the second summand leads to 
\begin{equation}
\label{eq-Khatform}
g(\widehat{H}(\lambda))
\;=\;
\frac{-1}{2\pi}\;
\int_{\RM^2}
dx\,dy\;
\partial_{\overline{z}}\widetilde{g}(x,y)\;
\frac{1}{z-\widehat{H}(\lambda)}
\,(\widehat{H}(\lambda)\,\Pi^*-\Pi\, H(\lambda))\,\frac{1}{z-H(\lambda)}\,\Pi^*
\;.
\end{equation}
Let us first check that this positive operator is indeed $\widehat{\Tt}$-traceclass by bounding its matrix elements:
$$
|\langle 0,n_2|g(\widehat{H}(\lambda))|0,n_2\rangle|
\,\leq \,
\int_{\RM^2}
\frac{dx\,dy}{2\pi}\;
|\partial_{\overline{z}}\widetilde{g}(x,y)|
\,
|\langle 0,n_2|
\frac{1}{z-\widehat{H}(\lambda)}
\,(\widehat{H}(\lambda)\,\Pi^*-\Pi\, H(\lambda))\,\frac{1}{z-H(\lambda)}
|0,n_2\rangle|
.
$$
Now we need to control the decay of the matrix elements of the resolvents of $H(\lambda)$ and $\widehat{H}(\lambda)$. This can be done by a Combes-Thomas estimate which states that there is an $\eta>0$ and $C_1$ such that
$$
|\langle n|(z-H(\lambda))^{-1}|m\rangle|
\;\leq\;
\frac{C_1}{|y|}\;e^{-\eta|y| |n-m|}
\;.
$$
Replacing this estimate and the corresponding one for the resolvent of $\widehat{H}(\lambda)$ shows that the matrix element $|\langle 0,n_2|g(\widehat{H}(\lambda))|0,n_2\rangle|$ is bounded above by a constant times
$$
\int_{\RM^2}
dx\,dy\;
|\partial_{\overline{z}}\widetilde{g}(x,y)|\,
\frac{1}{y^2}
\;
\sum_{m\in\ZM\times\NM}\,\sum_{k\in\ZM^2}\,
e^{-\eta|y| |n_2-m_2|-\eta |y| m_1}\,
|\langle m|(\widehat{H}(\lambda)\,\Pi^*-\Pi\, H(\lambda))|k\rangle|
\,e^{-\eta|y| |n_2-k_2|-\eta|y| |k_1|}
\;.
$$
%
Now the operator $H$ has only finite range hopping so that $\widehat{H}(\lambda)\,\Pi^*-\Pi\, H(\lambda)$ has non-vanishing matrix elements only in a finite distance away from the boundary (thus $m_2$ and $k_2$ are close to $0$). Also $m_1$ and $k_1$ are only a finite distance apart.  This shows
$$
\sum_{n_2\geq 0}\;
|\langle 0,n_2|g(\widehat{H}(\lambda))|0,n_2\rangle|
\;\leq\;
C_2\;
\int_{\RM^2}
dx\,dy\;
|\partial_{\overline{z}}\widetilde{g}(x,y)|\,
\frac{1}{y^2}
\;
\sum_{n_2\geq 0}\,\sum_{m_1\in\ZM}\,
e^{-\eta|y| |n_2|-\eta |y| m_1}\,
\;.
$$
Finally invoking the bound \eqref{eq-gtildebound} with $N=4$ and the fact that the integration domain is compact, one concludes that
$$
\widehat{\Tt}\bigl(g(\widehat{H}(\lambda))\bigr)
\;\leq\;
\sum_{n_2\geq 0}\;|\langle 0,n_2|g(\widehat{H}(\lambda))|0,n_2\rangle|
\;\leq\;C_3\,\|g\|_5
\;.
$$

\vspace{.2cm}

Next let us come to the perturbation theory in $\lambda$. It is based on
\begin{equation}
\label{eq-perturbsummand}
\widehat{H}(\lambda)\,\Pi^*-\Pi\, H(\lambda)
\;=\;
(\widehat{H}_0\,\Pi^*-\Pi\, H_0)\;+\;\lambda\,(\widehat{H}_1\,\Pi^*-\Pi\, H_1)
\;,
\end{equation}
and the resolvent identity
$$
\frac{1}{z-H(\lambda)}
\;=\;
\frac{1}{z-H_0}
\;+\;
\lambda\;\frac{1}{z-H_0}\;H_1\;\frac{1}{z-H(\lambda)}
\;,
$$
as well as the resolvent identity for $(z-\widehat{H}(\lambda))^{-1}$. Comparing with \eqref{eq-gdecomp} this leads to
\begin{eqnarray*}
G(\lambda) & = &
\frac{-1}{2\pi}\;
\int_{\RM^2}
dx\,dy\;
\partial_{\overline{z}}\widetilde{g}(x,y)\;
\left[\;\frac{1}{z-\widehat{H}_0}
\,H_1\,
\frac{1}{z-\widehat{H}(\lambda)}
\,(\widehat{H}(\lambda)\,\Pi^*-\Pi\, H(\lambda))\,\frac{1}{z-H(\lambda)}\,\Pi^*
\right.
\\
& & \hspace{4.4cm} +\;
\frac{1}{z-\widehat{H}_0}
\,(\widehat{H}_1\,\Pi^*-\Pi\, H_1)\,\frac{1}{z-H(\lambda)}\,\Pi^*
\\ 
& & 
\hspace{4.4cm} +\;
\left.
\frac{1}{z-\widehat{H}_0}
\,(\widehat{H}_0\,\Pi^*-\Pi\, H_0)\,\frac{1}{z-H_0}\,H_1\,\,\frac{1}{z-H(\lambda)}\,\Pi^*
\,\right]
\;.
\end{eqnarray*}
Because both $g(\widehat{H}(\lambda))$ and $g(\widehat{H}(0))$ are $\widehat{\Tt}$-traceclass due to the above, so is $G(\lambda)$. It follows that also $\widehat{J}_1(\lambda)G(\lambda)$ is $\widehat{\Tt}$-traceclass. Its trace will be bounded above again by a bound on its matrix elements. First of all, each of the summands contains $H_1$ or $\widehat{H}_1$ as a factor. Furthermore, each of the three summands contains one of the summands of \eqref{eq-perturbsummand} each of which is supported on a strip of finite width along the boundary. Combining this again with the Combes-Thomas estimates shows after similar estimates as above (with one supplementary factor $|y|^{-1}$ due to the extra resolvent)
$$
\left|\,\widehat{\Tt}(\widehat{J}_1(\lambda)G(\lambda))\,\right|
\;\leq\;
\sum_{n_2\geq 0}\,|\langle 0,n_2|\widehat{J}_1(\lambda)G(\lambda)|0,n_2\rangle|
\;\leq\;
C_4\,\|g\|_6\,\|H_1\|
\;.
$$
Hence it only remains to prove the bound 
$$
\left|\,\widehat{\Tt}\bigl((s^z\, [\widehat{H}_1,X_1]+[\widehat{H}_1,X_1]\,s^z) \,g(\widehat{H}_0)\bigr)\,\right|
\;\leq\;
C_5\,\|g\|_6\,\|H_1\|\;.
$$ 
This follows again by proving an upper bound on the matrix elements (using the fact that $\widehat{H}_1$ is only finite range so that the unboundedness of $X_1$ plays no role). This finishes the proof of the theorem.

\vspace{.2cm}

Let us conclude with a few technical remarks. By pushing the expansions above a bit further, one can calculate the linear growth in \eqref{eq-spinedgecur}, namely the coefficient $R_0(g)$ in $R(g,\lambda)=R_0(g)+\Oo(\lambda)$. Actually, one can control errors to arbitrary order and calculate a formal expansion in $\lambda$. Furthermore, in the situation where $g(H(\lambda))\not = 0$, the above calculation can still be carried through and edge currents can still be calculated by setting 
\begin{equation}
\label{eq-currentrenorm}
\widehat{\Tt}(\Pi g(H(\lambda))\Pi^* \widehat{J}_1(\lambda))\;=\;0
\;.
\end{equation}
This is reasonable because for $H(\lambda)$ on $\ell^2(\ZM^2)\otimes\CM^L$ the current $\EE_\PP(\langle 0|g(H(\lambda))J_1(\lambda)|0\rangle)$ in equilibrium indeed vanishes \cite{BES} and thus \eqref{eq-currentrenorm} just results by imposing that the expectation $\EE_\PP$ is calculated before the sum over $n_2$ in the trace $\widehat{\Tt}$. The remaining other term is then bounded by the above arguments. This procedure is similar to the one in \cite{EGS}.

\vspace{.5cm}

\noindent {\bf Note added in proof.} Recent experimental observations of \cite{DKSD} show that edge currents in QSH systems remain stable in relatively strong magnetic fields. This further supports the physical discussion in the introduction.



\end{document}